\documentclass[sigconf]{acmart}
\usepackage{fancyhdr}

\usepackage{booktabs} 

\usepackage{xspace}
\newcommand*{\eg}{e.g.\@\xspace}
\newcommand*{\ie}{i.e.\@\xspace}

\usepackage{graphicx}
\usepackage{caption}
\usepackage{subcaption}
\usepackage{multirow}
\usepackage{array}
\newcolumntype{L}[1]{>{\raggedright\let\newline\\\arraybackslash\hspace{0pt}}m{#1}}
\newcolumntype{C}[1]{>{\centering\let\newline\\\arraybackslash\hspace{0pt}}m{#1}}
\newcolumntype{R}[1]{>{\raggedleft\let\newline\\\arraybackslash\hspace{0pt}}m{#1}}

\usepackage{hyperref}

\fancypagestyle{firststyle}
{
   \fancyhf{}
   \fancyhead[C]{To appear in Proceedings of the 25th ACM International Conference on Multimedia 2017}
   \fancyfoot[C]{}
}

\begin{document}

\copyrightyear{2017} 
\acmYear{2017} 
\setcopyright{acmlicensed}
\acmConference{MM '17}{October 23--27, 2017}{Mountain View, CA, USA}\acmPrice{15.00}\acmDOI{10.1145/3123266.3123282}
\acmISBN{978-1-4503-4906-2/17/10}

\title{Protest Activity Detection and Perceived Violence Estimation \\ from Social Media Images}

\author{Donghyeon Won}
 \affiliation{%
   \institution{UCLA} 
 }
 \email{dh.won@ucla.edu}

\author{Zachary C. Steinert-Threlkeld}
 \affiliation{%
   \institution{UCLA} 
 }
 \email{zst@luskin.ucla.edu}

\author{Jungseock Joo}
 \affiliation{%
   \institution{UCLA} 
 }
 \email{jjoo@comm.ucla.edu}

\begin{abstract}
We develop a novel visual model which can recognize protesters,  describe their activities by visual attributes and estimate the level of perceived violence in an image. Studies of social media and protests use natural language processing to track how individuals use hashtags and links, often with a focus on those items' diffusion.  These approaches, however, may not be effective in fully characterizing actual real-world protests (\eg, violent or peaceful) or estimating the demographics of participants (\eg, age, gender, and race) and their emotions. Our system characterizes protests along these dimensions. We have collected geotagged tweets and their images from 2013-2017 and analyzed multiple major protest events in that period. A multi-task convolutional neural network is employed in order to automatically classify the presence of protesters in an image and predict its visual attributes, perceived violence and exhibited emotions. We also release the \textbf{UCLA Protest Image Dataset}, our novel dataset of 40,764 images (11,659 protest images and hard negatives) with various annotations of visual attributes and sentiments. Using this dataset, we train our model and demonstrate its effectiveness. We also present experimental results from various analysis on geotagged image data in several prevalent protest events. Our dataset will be made accessible at 
\href{https://www.sscnet.ucla.edu/comm/jjoo/mm-protest/}{https://www.sscnet.ucla.edu/comm/jjoo/mm-protest/}.
\end{abstract}

%
%

\ccsdesc[500]{Information systems~Multimedia information systems}
\ccsdesc[500]{Computing methodologies~Computer vision}
\ccsdesc[500]{Computing methodologies~Activity recognition and understanding}
\ccsdesc[300]{Computing methodologies~Scene understanding}
\keywords{Protest; Action and Activity Recognition; Scene Understanding; Social Media Analysis; Visual Sentiment Analysis}

\maketitle

\thispagestyle{firststyle}

\section{Introduction: \\ Images and Political Contention}

Online social media have served as an open information channel which hosts public discussions on a number of social and political issues. Individuals respond to real world events in social media, and their responses can influence various social events and provoke public movements. As an alternative to traditional mass media outlets, social media, its users, and the content shared on them are often independent of government authorities. In particular, the possible impact of social media on \textit{protests} has been analyzed in the context of the Arab Spring \cite{Steinert-Threlkeld2015a,Steinert-Threlkeld2017}, social movements in Europe \cite{Gonzalez-Bailon2013}, and election protests in Russia \cite{Enikolopov2016}.

Social scientists have long studied protests, but the difficulty of acquiring and processing large-scale data has limited what questions are answerable.  Work in this field is therefore dominated by formal and qualitative models \cite{Kuran1989,Lohmann1994, McAdam1986,Tullock1971} or surveys of protest participants after a protest occurs \cite{Muller1986,Tufekci2014}.   The spread of information and communications technologies (ICT), like the Internet and cell phones, has generated a new burst of theorizing and testing, with scholars modeling whether these technologies lead to more protest \cite{Little2015,Shapiro2015} and using data on hundreds of thousands to millions of people across countries and times to test these models \cite{Barbera2015c,Driscoll2017, Gonzalez-Bailon2013,Steinert-Threlkeld2017}.  

With such data, scholars can now measure the behavior of various people across multiple cities and countries for weeks, months, or years.	The advantage of these data has been that scholars can see what protesters say.  In the last few years, however, accounts have started to share \textbf{images} with greater frequency, and scholars have yet to analyze what protesters show. 
    
    The objective of this paper is to develop an automated system that analyzes what images are shared during protests and how they change over space and time. In doing so, our visual approach identifies salient characteristics of protests -- especially \textbf{violence} -- by automatically assessing visual activities and attributes described in a given scene. 
    
    Violence is a critical dimension of protest in understanding social mobilization, as violent protests typically generate a much higher level of media and public attention. There might be other cues that one could use to approximate the level of violence in a protest, such as police or government statements or the number of people who have been killed, injured, or arrested. However, this information can be often inaccurate or not provided at all to the public in an official channel. Therefore, the goal of our study is to take advantage of unfiltered stream of data in social media and to assess the level of perceived violence for protest events.

\begin{figure*}
\centering
\includegraphics[width=1\textwidth]{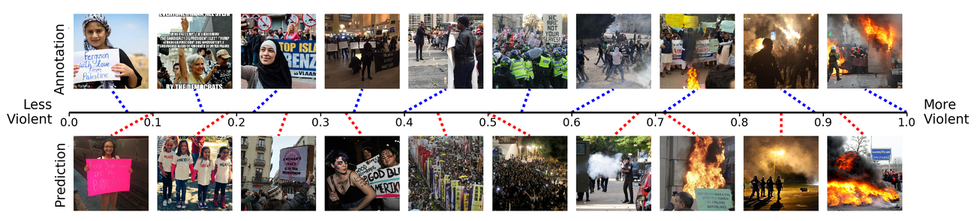}
\caption{Sample images in our Protest Image dataset ordered by their perceived violence scores: (top) annotation (bottom) prediction. }
\label{fig:violence-one-d}
\end{figure*}

    The key contributions of our paper can be summarized as follows. 
    
\begin{itemize}
  \item We have collected, and will release, a novel dataset of protest images with human coded data of perceived violence and image sentiments. Our dataset is an order of magnitude larger than existing social event image datasets and provides fine-grained annotations which are directly relevant to protest research.  
  \item We train a model based on a Convolutional Neural Network (CNN) that can automatically classify the content of images, especially perceived violence and sentiments, all of which are inferred jointly from the shared visual feature representation. 
  \item We have collected geotagged tweets and associated images across the world from August 2013 to detect, track, and analyze various protests in different countries. In this paper, we analyze and compare five protest events including Black Lives Matter and Women's March. Our analysis reveals that the degree of predicted perceived violence differs significantly across events and also across states within an event. 
\end{itemize}

    


\section{Related Work}


Although understanding protest and violence has been a critical topic of research in political science, there are few work in the fields of political science and media studies which attempts to automatically analyze visual or multimodal data due to the lack of proper methods and datasets.    

    While this paper is the first work that analyzes what images are shared during protests and how those change across events, recent studies in \textbf{``social'' multimedia and computer vision} employ large-scale visual content analysis to tackle related research questions in political science, media studies and communication. For instance, facial attribute classification (gender, race, and age; \cite{liu2015deep}) has been used to examine the supporter demographics of major politicians in the U.S. using profile photographs of Twitter users \cite{wang2016deciphering}. Researchers have also analyzed photographs of politicians shared on social media \cite{you2015multifaceted} or their perceived personalities from facial appearance \cite{joo2015automated}. Public opinion about politicians has been also studied in relation to visual portrayals and persuasions in mass media \cite{Joo2014} and presentations in social media \cite{Anastasopoulos2016}. These works all highlight the importance of the visual cue in human perception of media content and the advances in multimedia and computer vision have enabled to recognize subjective, perceived dimensions of images or videos, \eg, creative \cite{redi20146}, interesting \cite{grabner2013visual}, or sexually provocative \cite{ganguly2017detecting}.
    
    
    Among more traditional work in \textbf{multimedia}, our paper is closely related to social event detection or classification \cite{brenner2012social,reuter2013social,petkos2012social,petkos2014graph,yang2015cross,qian2015social}. While these studies focus on identifying the same type of event (\ie, clustering) or classifying event type, we specifically concentrate on the protest event and investigate various ways to characterize them. 

In addition, there have been a few works which propose to automatically classify violent activities from images or videos by static image or motion features \cite{nievas2011violence, de2010violence, chen2011violence, hassner2012violent}. These existing studies on violence detection are mostly concerned with physical violence such as physical fights between players in sports games \cite{nievas2011violence}, detecting bloody frames in movies \cite{chen2011violence}, or aggressive behavior in short videos \cite{de2010violence}. Our work clearly differs from these works as we focus on perceived violence in protest activities of various types which include not only physical assaults, but also rallies, demonstrations, or even peaceful gatherings.

\section{UCLA Protest Image Dataset}

In this section, we describe our dataset of social protest images. It is designed to support studies in protest activity detection, fine-grained attribute recognition, and visual sentiment analysis.\footnote{\href{https://www.sscnet.ucla.edu/comm/jjoo/mm-protest/}{https://www.sscnet.ucla.edu/comm/jjoo/mm-protest/}.}

Prior studies have proposed image datasets for general social event detection such as music concerts or birthday parties. However, our main research topic, protest and public mobilization, has not been sufficiently addressed in these datasets because they lack other rich annotations. The social event detection benchmark \cite{reuter2013social} has released an image dataset containing images of the protest category; however, a very small portion of the dataset was composed of protest images (800-1000 images) and the images do not have any other annotations than categorical information. Thus, this dataset is insufficient to conduct in-depth studies specifically aimed at analysis of protest images. 
%
%


\begin{figure}
\centering
\includegraphics[width=0.47\textwidth]{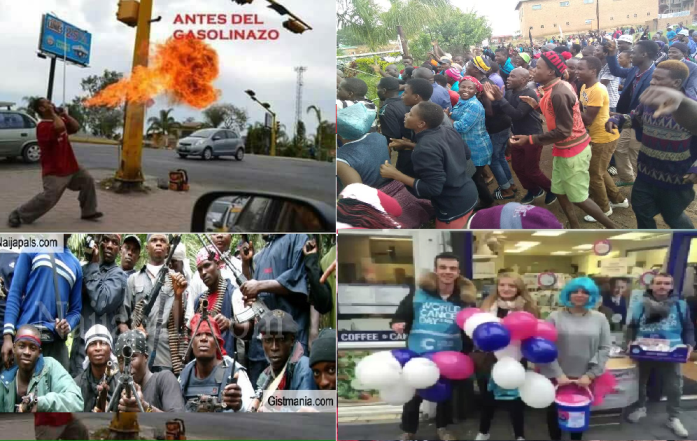}
\caption{Hard negative examples (non-protest) in our dataset. The images exhibit visual features common in protest images such as fire, a group of people, or a weapon.}
\label{fig:data_prot}
\end{figure}

Our dataset contains 40,764 images which have been collected from Twitter and other online resources. 
11,659 images are protest images identified by annotators and the rest are hard-negative images (\eg, crowd in stadium). A few negative examples are shown in Fig.~\ref{fig:data_prot} and positive examples with annotations and prediction scores of violence are presented in Fig.~\ref{fig:violence-one-d} and Fig.~\ref{fig:sample}. Each positive protest image was annotated for its visual attributes (\eg, children, fire, or large crowd) and image sentiment. 

\subsection{Image Collection}
%
%

Our model should be able to distinguish between a protest crowd and other large gathering such as concerts or sporting events.  It should also distinguish between non-violent and violent protests.  In order to effectively capture diverse visual patterns of protests and train a robust model, we collect images from multiple sources by web search (Google Image search) and also our own Twitter data stream in a refining, active learning approach. 

We first collected 10,000 images which may describe a protest scene by web search using a set of keywords. We manually selected some general keywords (\eg, ``protest'', ``riot'', ``demonstration'') and also used the names of recent major protest events (\eg, ``Black Lives Matter'' or ``Women's March'') based on the Wikipedia page of the list of  protests.\footnote{\url{http://en.wikipedia.org/wiki/List_of_riots}} \footnote{\url{https://en.wikipedia.org/wiki/List_of_ongoing_protests}} 
We trained our first, rough classification CNN, treating these noisy images as positive examples. For negative examples, we used keywords such as ``concert'' or ``stadium.'' Our model architecture is based on a 50-layer ResNet \cite{he2016deep} (See Sec. \ref{models}). This model was applied back to the initial image set to filter out images of very low scores (\ie, easy negative) as they are unlikely to be protest scenes. 


We then applied this classifier to random samples from our geotagged Twitter image collection (see Sec. \ref{geo-coded} for detail), without any filtering by keywords, and obtained a set of images whose prediction scores were above a threshold. The optimal threshold was selected empirically from a set of hand-labeled images such that it prunes the majority of irrelevant images while keeping most of the positive examples. Therefore, this set contains many hard negative examples, such as flash mobs. These two sets of images were then merged and provided to the annotators from Amazon Mechanical Turk who labeled the presence or absence of a protester in each image. 

%
%

\subsection{Violence and Emotions in Protest Images}
Each protest event is attended by different people who gather together for different purposes \cite{Fisher2014}. The degree of violence involved in a protest can also vary greatly with the participants and their demands \cite{Feinberga2017}. Publicly shared photographs allow us to assess how a protest event is depicted: how \textbf{violent} it is and what kinds of \textbf{emotions} are expressed. 

\begin{figure}
\centering
\includegraphics[width=0.23\textwidth]{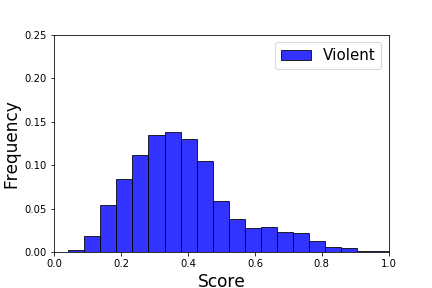} \\
\includegraphics[width=0.23\textwidth]{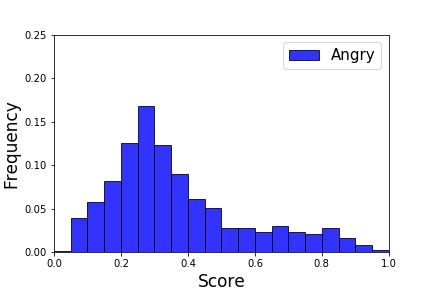}
\includegraphics[width=0.23\textwidth]{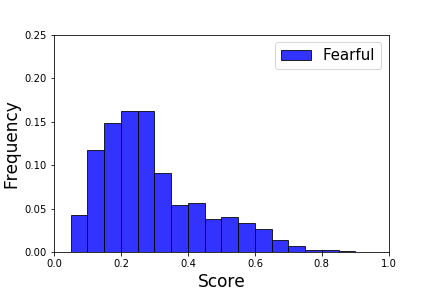}
\includegraphics[width=0.23\textwidth]{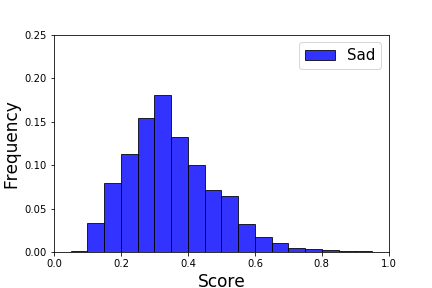}
\includegraphics[width=0.23\textwidth]{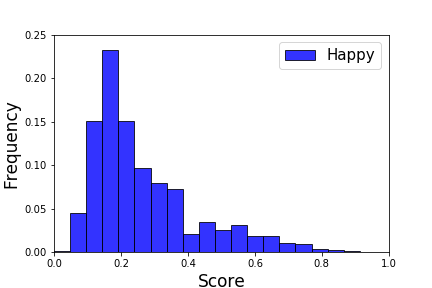}
\caption{Distributions of perceived violence and image sentiment scores rated by annotators.}
\label{fig:anno_hist}
\end{figure}

Many emotional dimensions, such as anger or fear, are highly correlated with perceived violence, but they sometimes capture different traits. The distributions of annotations also differ across dimensions, as shown in Fig. ~\ref{fig:anno_hist}. For instance, protesters might be angry but still not violent. To reduce the annotation cost, we excluded the two emotional dimensions of surprise and disgust as they overlap with other dimensions. 


In addition, we further identified common scene attributes associated with protest images and annotated them in order to analyze what kinds of visual attributes are correlated with the perceived violence or image sentiments. We first generated any related visual concepts for each image in a small subset of the image set and constructed the most common attributes among them which are shown in Table ~\ref{table:list-visual-attribute}. 


\subsection{Image Annotation}
We used Amazon Mechanical Turk (AMT) to obtain necessary annotations for each image in our dataset including (1) whether an image contains a protest activity or protesters, (2) visual attributes in the scene, (3) the level of perceived violence and other sentiments. The first two tasks require objective, binary annotations. We assigned two workers to each image for these tasks and confirmed the values when both workers agreed (if not, the image was sent to the third judge). 

On the other hand, since our perceived violence is a subjective and continuous variable, we instead requested pairwise comparison-based annotations \cite{parikh2011relative}. Specifically, we randomly sampled 58,295 image pairs among 11,659 protest images such that each image is paired up 10 times and assigned 10
workers to evaluate each pair. For each pair, the annotators were asked to choose an image which looks more violent than the other. We used the Bradley-Terry model \cite{bradley1952rank} and estimated the global scores for images such that each image is assigned a real-number score of perceived violence. 

The advantages of pairwise annotation method has been well understood in prior work \cite{parikh2011relative,kovashka2012whittlesearch}, but it requires much more annotations to be collected. To ease the burden of the overall annotation task, the remaining emotional sentiment annotations (angry, fearful, sad, happy) were obtained by individual evaluation (\ie, the annotator was given only one image at a time and asked to provide his response.) In both cases (violence and emotions), we obtained a scalar value in $[0, 1]$.

%
%
%



\begin{table}
\caption{List of visual attributes.}
\label{table:list-visual-attribute}
\begin{center}
  \begin{tabular}{ |C{1.3cm} | C{6.2cm}|}
  \hline
  Attribute & Description \\ \hline \hline
  Sign &  A protester holding a visual sign (on paper, panel, or wood). \\ \hline
  Photo &  A protester holding a sign containing a photograph of a person (politicians or celebrities) \\ \hline
  Fire & There is fire or smoke in the scene. \\ \hline
  Law enf. & Police or troops are present in the scene. \\ \hline
  Children & Children are in the scene. \\ \hline
  Group 20 & There are roughly more than 20 people in the scene. \\ \hline
  Group 100 & There are roughly more than 100 people in the scene. \\ \hline
  Flag & There are flags in the scene \\ \hline
  Night & It is at night. \\ \hline
  Shout & One or more people shouting. \\
     \hline      
  \end{tabular}
\end{center}
\end{table}


  

\section{Models}
\label{models}

\begin{figure*}
\centering
\includegraphics[width=0.24\textwidth]{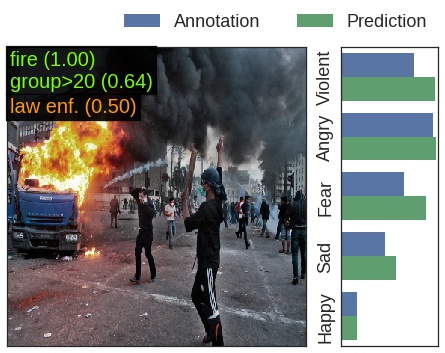}
\includegraphics[width=0.24\textwidth]{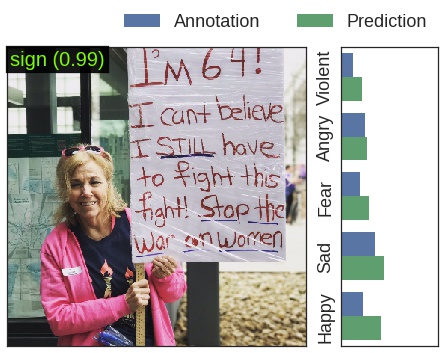}
\includegraphics[width=0.24\textwidth]{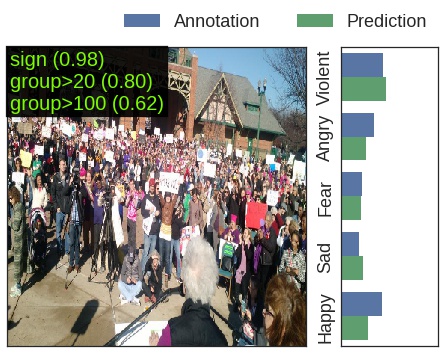}
\includegraphics[width=0.24\textwidth]{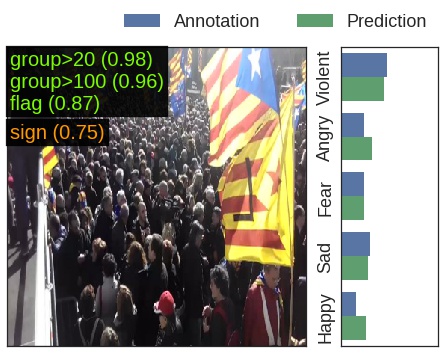}
\includegraphics[width=0.24\textwidth]{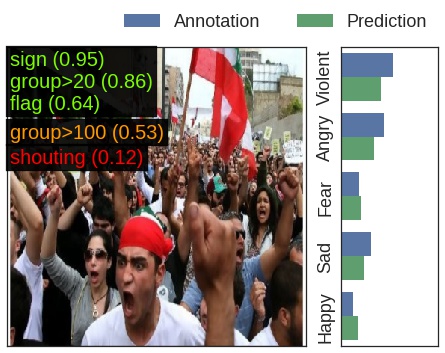}
\includegraphics[width=0.24\textwidth]{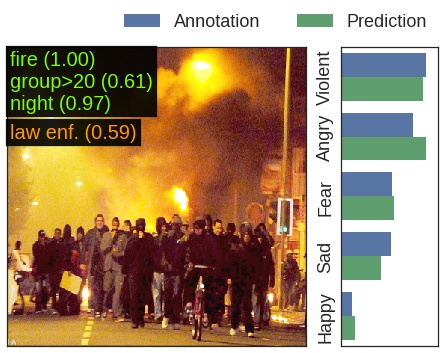}
\includegraphics[width=0.24\textwidth]{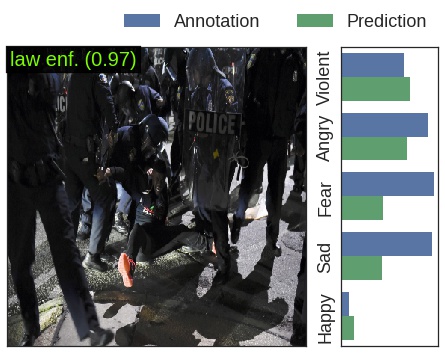}
\includegraphics[width=0.24\textwidth]{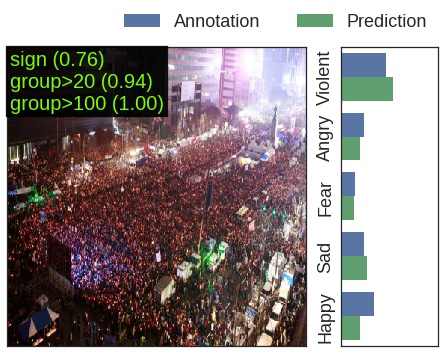}
\includegraphics[width=0.24\textwidth]{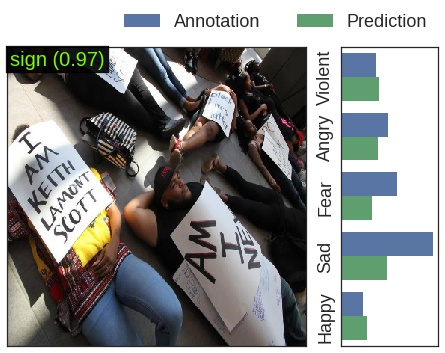}
\includegraphics[width=0.24\textwidth]{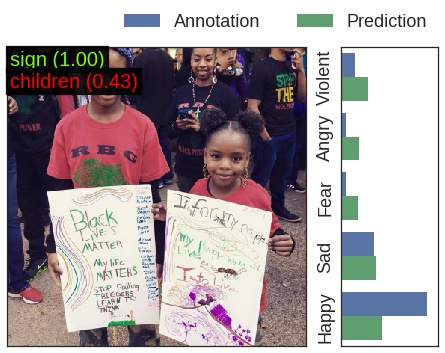}
\includegraphics[width=0.24\textwidth]{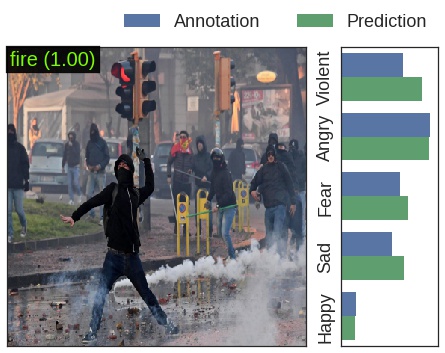}
\includegraphics[width=0.24\textwidth]{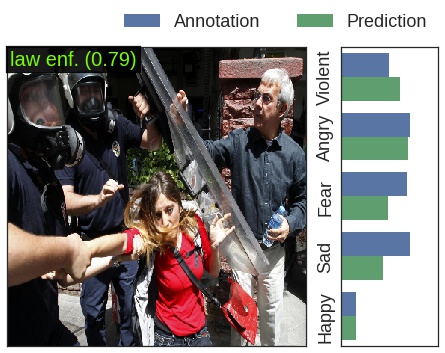}
\caption{Example images in our protest dataset with various scores obtained by our model: perceived violence, sentiments, and visual attributes. }
\label{fig:sample}
\end{figure*}


We use two separate models to recognize protest activities in images. First, we train a CNN which takes a full image as input and outputs a series of prediction scores including the binary image category (\ie, protest or non-protest) (1), visual attributes (10), and perceived violence and image sentiment ($1+4$). Our model architecture is based on a 50-layer ResNet \cite{he2016deep}, 
  which consists of 50 convolutional layers with batch normalization and ReLU layers. The architecture of the model is briefly described in Table \ref{table:resnet}. The features computed through convolutional layers are all shared by linear layers for multiple classification tasks. 
We jointly train the model such that all parameters for 3 different tasks -- protest classification, violence and sentiment estimation, and visual attribute classification -- are updated jointly. We use binary cross entropy loss to train our binary variables (protest and visual attributes) and mean squared error to train violence and sentiment dimensions.

\begin{table}\footnotesize
  \caption{ The architecture of our model.}
  \label{table:resnet}
  \begin{center}
  \begin{tabular}{|c|c|C{0.8cm}|C{0.8cm}|C{1cm}|C{1cm}|} 
  \hline
  Layer & Output size & \multicolumn{4}{c|}{ Building blocks} \\ \hline
  conv1 & $112 \times 112$ & \multicolumn{4}{c|}{$7\times7, 64,$ stride $2$} \\ \hline
  \multirow{4}{*}{ conv2 } & \multirow{4}{*}{ $56 \times 56$ } & \multicolumn{4}{c|}{$3\times3$ max pool, stride 2 }\\ 
  \cline{3-6}
  && \multicolumn{4}{c|}{$\left[\begin{array}{ccc} 1\times1 ,& 64  \\ 3\times3,& 64 \\ 1\times1,& 256 \\ \end{array}\right]\times3$} \\ \hline
  conv3 & $28 \times 28$ & \multicolumn{4}{c|}{$\left[\begin{array}{ccc} 1\times1 ,& 128  \\ 3\times3,& 128 \\ 1\times1,& 512 \\ \end{array}\right]\times4$} \\ \hline
  conv4 & $14 \times 14$ & \multicolumn{4}{c|}{$\left[\begin{array}{ccc} 1\times1 ,& 256  \\ 3\times3,& 256 \\ 1\times1,& 1024 \\ \end{array}\right]\times6$} \\ \hline
  conv5 & $7 \times 7$ & \multicolumn{4}{c|}{$\left[\begin{array}{ccc} 1\times1 ,& 512  \\ 3\times3,& 512 \\ 1\times1,& 2048 \\ \end{array}\right]\times3$} \\ \hline
  pooling &2048&\multicolumn{4}{c|}{average pooling} \\ \hline
 classification   & $17$ & 1-d fc (protest) &1-d fc (violence) &4-d fc (sentiment) &10-d fc (visual attribute)\\
     \hline      
  \end{tabular}
\end{center}
\end{table}

In addition, another CNN captures various facial attributes from images. We use OpenFace \cite{amos2016openface} for our face model, which was developed for face recognition. We use the CelebA facial attribute dataset to train the attribute model.  That model outputs gender, race, and other expressions \cite{liu2015deep}. For each image, we use dlib's face detection and alignment\footnote{dlib.net} and crop the internal facial region to feed into the facial CNN model.
In our analysis, facial attributes are especially important because social scientists have theorized about the role of emotions in leading to and sustaining protests, but arguments have had to rely on qualitative models and case studies \cite{Yang2000,Mercer2010,Pearlman2013}.  With our model, we can now test these theories with more precision than before.

\section{Results}
This section presents various experimental results obtained from our analysis. We discuss the general performance of our model and then provide the results of actual analyses conducted on our geocoded tweet dataset of protests. 

\subsection{Inter-rater reliability}

\begin{center}
\begin{table}
  \caption{Inter-rater reliability measured by Pearson's correlation coefficients between two randomly split annotator groups. }
  \label{table:dataset_interanno}
  \begin{tabular}{ c | C{1cm}  C{1cm}  C{1cm}  C{1cm}  C{1cm}  }
              & Violent & Angry & Fearful & Sad & Happy \\ \hline
     Pearson's $\rho$ & .716 & .568 & .417 & .362 &.875 \\ \hline
  \end{tabular}
\end{table}
\end{center}


Table~\ref{table:dataset_interanno} reports the Pearson correlation coefficient for inter-rater reliability between two randomly split annotator groups. As we used Mechanical Turk to collect annotations, it is inappropriate to apply a standard test which typically assumes complete data. Therefore, we measure correlations between non-overlapping groups of workers. This method has also been frequently used in the literature \cite{isola2011makes}. The results are all statistically significant. 

\begin{table}
\begin{center}
  \caption{Pearson's correlation coefficients between visual sentiments and visual attributes, measured from annotations. We only print fields which are statistically significant (p-val $<$ 0.0001).}
  \label{table:sent_dim_corr}
  \begin{tabular}{ c | C{1cm} C{1cm}  C{1cm}  C{1cm}  C{1cm}  C{1cm}  }
    & Violent & Angry & Fearful & Sad & Happy \\ \hline \hline
    Violent &  & .671 & .575 & .351 & -.359   \\
    Angry & &  & .795 & .626 & -.427   \\ 
    Fearful & &  &  & .752 & -.219   \\ 
    Sad &  & &  &  & -.195   \\ 
    \hline
    Sign  & -.479 & -.549 & -.495 & -.288 & .225   \\ 
    Photo & -.047 &  &  &  &    \\ 
    Fire  & .567 & .578 & .504 & .297 & -.184   \\ 
    Law enf.  & .367  & .417 & .399  & .239  & -.186  \\ 
    Group $>$100 & .152 & -.166  & -.279 & -.147 &  \\ 
    Night  & .206 & .183  & .143 & .086 & -.129 \\ 
    Shouting &  & .106 &  &  & -.087  \\ 
    \hline
  \end{tabular}
\end{center}
\end{table}
\begin{figure}
\centering
\includegraphics[width=0.22\textwidth]{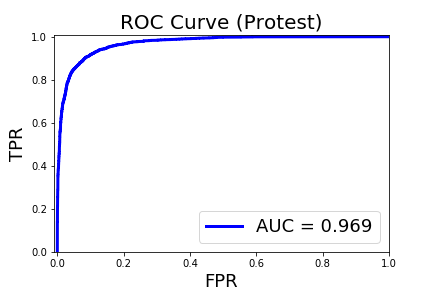}
\includegraphics[width=0.22\textwidth]{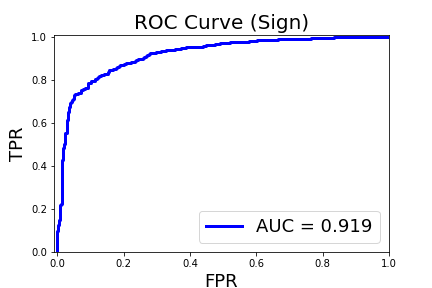}
\includegraphics[width=0.22\textwidth]{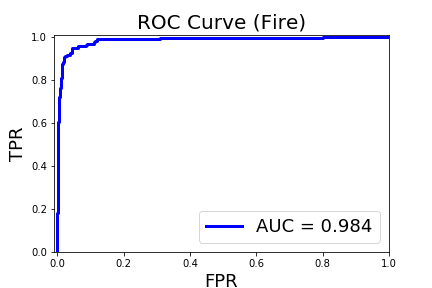}
\includegraphics[width=0.22\textwidth]{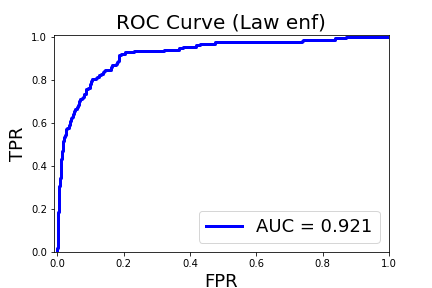}
\includegraphics[width=0.22\textwidth]{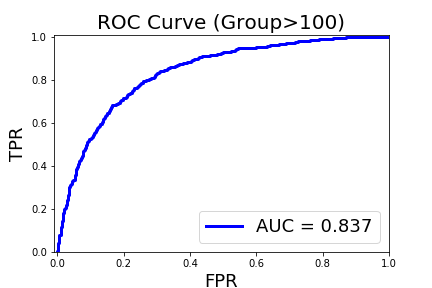}
\includegraphics[width=0.22\textwidth]{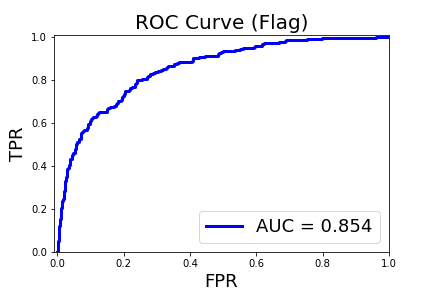}
\includegraphics[width=0.22\textwidth]{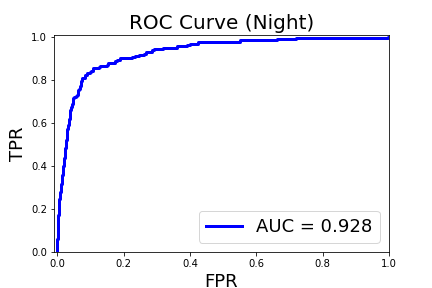}
\includegraphics[width=0.22\textwidth]{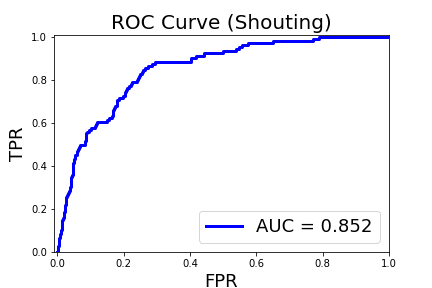}
\caption{ROC curves for protest image and visual attribute classifications.}
\label{fig:roc}
\end{figure}

\begin{table}\footnotesize
\begin{center}
  \caption{Performance of protest scene and attributes classification. AUC is area-under-curve in a ROC curve.}
  \label{table:perform_class}
  \begin{tabular}{| c || c || c | c | c | c | c | }
	\hline  
    Fields & Protest & Sign & Photo & Fire & Law & Children \\ \hline
    Pos. rate & .286 & .829 & .036 & .057 & .067  & 0.030\\ \hline
    AUC & .969 & .919 & .738 & .984 & .921 & .813 \\ 
    \hline
    \hline
    Fields & $\cdot$ & Group $>20$  & Group$>100$ & Flag & Night & Shout \\ \hline
    Pos. rate & & .730 & .252 & .083 & .084 & .047 \\ \hline
    AUC & & .795 & .837 & .854 & .928 & .852\\ \hline
  \end{tabular}
\end{center}
\end{table}
\subsection{Model Performance}
\subsubsection{Protest Scene and Attributes Classification. }
We randomly split our entire dataset into a training set (80\%) and a test set (20\%). 
Table~\ref{table:perform_class} shows the classification accuracies for protest scene classification and visual attribute classification, measured on the test set. The ROC curves for selected variables are also shown in Fig.~\ref{fig:roc}. Some variables such as `children' or `shouting' only have a very small number of positive examples, but our model in general achieved reasonable accuracies for most variables.

\begin{table}
  \caption{Perceived violence and image sentiment prediction accuracy of our model measured by Pearson's correlation coefficients and $r^2$ values. }
\label{table:sent_accuracy}
\begin{center}
  \begin{tabular}{ c | C{1cm} C{1cm}  C{1cm}  C{1cm}  C{1cm} }
             & Violent & Angry & Fearful & Sad & Happy \\ \hline
     Pearson's $\rho$ & .900 & .753 & .626 & .340 & .382  \\ \hline
     $r^2$ &.809 &.566 &.392 &.116 &.146\\ 
     \hline
  \end{tabular}
\end{center}
\end{table}

\subsubsection{Violence and Sentiment Estimation}
\begin{figure}
\centering
\includegraphics[width=0.4\textwidth]{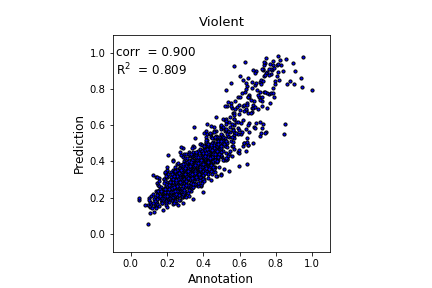}
\caption{A scatter plot of perceived violence in annotations and predictions.}
\label{fig:scatter_violent}
\end{figure}

We also measure the accuracy of our model in estimating perceived violence and image sentiments on emotional dimensions. Table~\ref{table:sent_accuracy} reports the Pearson's linear correlation coefficients and the coefficients of determination ($r^2$) between human annotations and our model's predictions, measured on the test set. Fig.~\ref{fig:scatter_violent} shows the scatter plot of annotations and predictions. 

We found that our model performs very well in predicting image violence. It is less accurate for emotional sentiments; we believe this is at least partly because the individual annotation scheme (vs. pairwise) sometimes led to less consistent annotations across annotators (\eg, due to the lack of a reference scale). 

Fig.~\ref{fig:sample} shows qualitative examples in our test dataset. While our model successfully predicts violence and image sentiment, there are some difficult cases where the prediction does not match annotators' ratings. We found the most important factor that our model does not address very well is a semantic relation between uncommon visual feature (symbolic gestures such as ``die-in'' in Black Lives Matter) and their meanings and associated emotions. For instance, in the bottom-left image, people demonstrate at the protest by pretending that they are causalities. However, the model might have treated this group gesture as people who are actually wounded, and some images in our dataset contain actually wounded people. 

\subsubsection{What makes a protest image violent?}
We now analyze visual attributes common in images which annotators rate as violent.  We identify these features by measuring correlations between visual attributes and perceived sentiments (from annotations). As shown in Table \ref{table:sent_dim_corr}, annotators find images with more dangerous (or potentially more dangerous) physical activities (`fire' and `law enforcement') as more violent, angry and fearful than images with a `big group' of people holding `signs.' 

\begin{center}
\begin{table}
  \caption{Correlation coefficients between facial attributes and perceived violence and image sentiment. (p-values $< 0.0001$) }
  \begin{tabular}{ c | C{1cm}  C{1cm}  C{1cm}  C{1cm}  C{1cm}  }
              & Violent & Angry & Fearful & Sad & Happy \\ \hline
     Male & .120 & .146 & .145 & .137 & -.151 \\     
     White & -.166 & -.172 & -.171 & -.160 & .158 \\ 
     Black & .189 & .193 & .194 & .180 & -.143 \\ 
     Smile & -.151 & -.196 & -.186 & -.145 & .229 \\ 
     Frown & .181 & .223 & .210 & .175 & -.237 \\  \hline
  \end{tabular}
\label{table:sent_face}
\end{table}
\end{center}

Another important set of visual features are human attributes, including expressions or demographic information about human subjects in a scene.  We assess these features from their faces. The correlations between the facial attributes and sentiment predictions are presented in Table.~\ref{table:sent_face}. 

From this analysis, we find smiling faces negatively correlate with perceived violence and other emotional dimensions such as angry or fearful (but not happy). Perceived violence also differs across gender and race groups. This could arise if some demographic subgroups are involved in a more violent protest activity captured in each photograph, or the level of violence varies in different underlying protest events which have different groups of participants (but the violence is exogenous to the participants). 
%
%

\subsection{Protest Event Analysis}
The key advantage of using photographs in our analysis is to assess various non-verbal characteristics of protest events. The following subsections present the results obtained by applying the trained model to the tweet data stream collected over the past three years.

\begin{figure}
\centering
\includegraphics[width=0.5\textwidth]{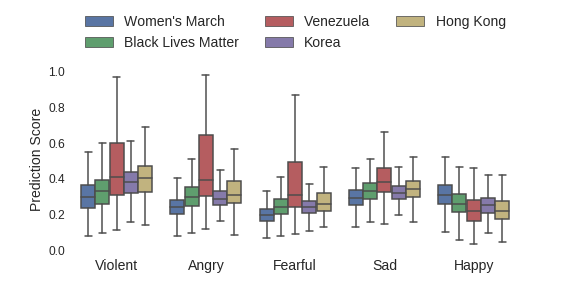}
\caption{Predicted violence and sentiments of tweet images in different protest events. The box represents the range between the first and third quartiles.}
\label{fig:event_senti}
\end{figure}

\begin{figure}
\centering
\includegraphics[width=0.5\textwidth]{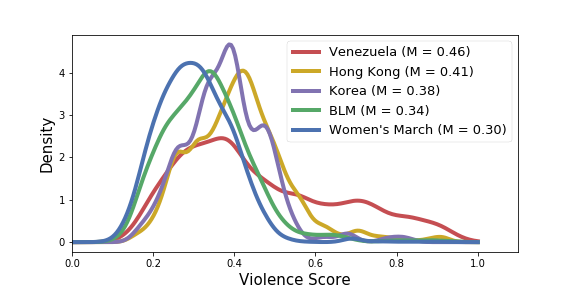}
\caption{Distribution of predicted violence scores of images in different protest events. }
\label{fig:violence-dist}
\end{figure}

\begin{figure}
\centering
\includegraphics[width=0.5\textwidth]{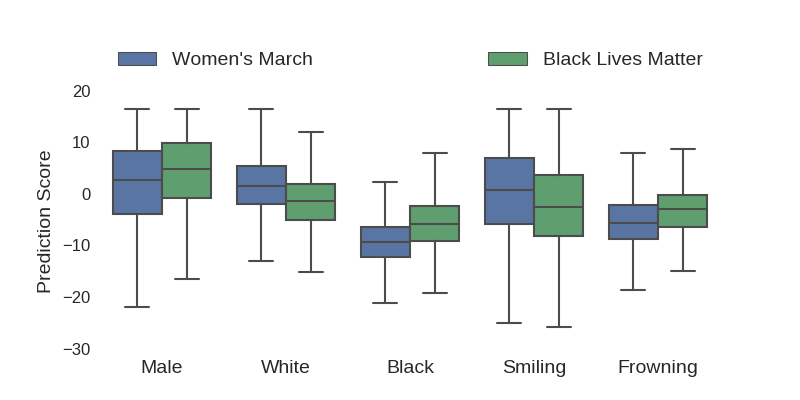}
\caption{Predicted face attributes in tweet images from the Women's March and Black Lives Matter. They are separate dimensions, so one cannot directly compare the absolute score of White and that of Black. The box represents the range between the first and third quartiles.}
\label{fig:event_face}
\end{figure}

\begin{figure*}
\centering
\includegraphics[width=0.92\textwidth]{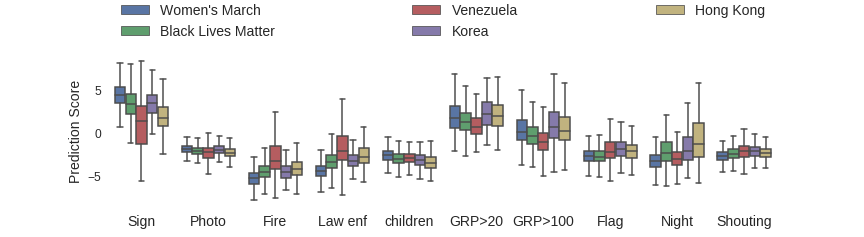}
\caption{Predicted visual attributes in tweet images in different protest events. The box represents the range between the First and Third quartiles.}
\label{fig:event_var}
\end{figure*}

Fig.~\ref{fig:event_senti} shows that our image analysis reveals two interesting results when comparing the Women's March, Black Lives Matter, protests in South Korea, Hong Kong, and Venezuela.\footnote{We obtained event specific tweets in the following way. For Black Lives Matter, which has spanned for 3 years across the country, we simply filtered tweets by the hashtag of \#BlackLivesMatter. For other events, we specified the region and date for which we know the protests happened and classified images to obtain protest related tweets.}  First, protests in Venezuela are more violent and angrier than the other protests; the standard errors are large, but the effect is persistent across the five emotions.  This result matches our prior beliefs, as Venezuelan protests, especially recently, have been violently repressed \cite{Lopez2017}.  Second, the Women's March is the least violent and angry, and this pattern holds across the five emotions as well. The statistics of these two events are statistically significant (p-val $<$ 0.00001) when compared to the other protests. The distributions of violence scores for each event are shown in Fig.~\ref{fig:violence-dist}. 

Fig.~\ref{fig:event_face} shows there were more female protesters in the Women's March protest and more African-Americans in the Black Lives Matter protests. We manually verified the gender ratio of these two events by counting randomly sampled 500 detected faces. In case of Women's March, the gender ratio between male and female was $0.22:0.78$. On the other hand, in Black Lives Matter images, the gender ratio was $0.46:0.54$. Recent studies have attempted to measure the demographics of social media users involved in a social movement by analyzing their profile photographs \cite{olteanu2016characterizing,an2016greysanatomy}. In contrast, our measures and data focus on the demographics of \textit{actual} protesters. 

Fig.~\ref{fig:event_var} shows the difference of predicted visual attributes in different events. Venezuelan protesters used less protest signs than protesters in other countries. The event that had most protest signs was Women's March. In contrast, fire was detected in Venezuelan protest images most frequently. Also, images related to law enforcements appeared most frequently in Venezuelan protests. These are all the indicators of a higher level of violence as discussed above. Another interesting result to note is that images with large groups appeared most frequently in Korean protest images since the recent Korean protests were very effectively organized on every Saturday for a few months.

\subsection{Geo-coded Tweet Analysis}
\label{geo-coded}
We have collected tweets from August 26, 2013 to the present, asking Twitter's streaming API for only tweets containing GPS coordinates.  Twitter returns all tweets with GPS coordinates up to 1\% of the total volume of tweets at a given time; since approximately 2\% of tweets contain GPS coordinates \cite{Leetaru2013a,Bastos2014}, we estimate we have collected half of all tweets with GPS coordinates (approximately 6.4 billion).   We therefore have precise location information for all tweets and their images in our dataset, allowing us to to track the spatial distribution of the protest event coverage in Twitter. 

\begin{figure}
    \centering
    \begin{subfigure}[b]{0.24\textwidth}
    \centering
		\includegraphics[height=1.1in]{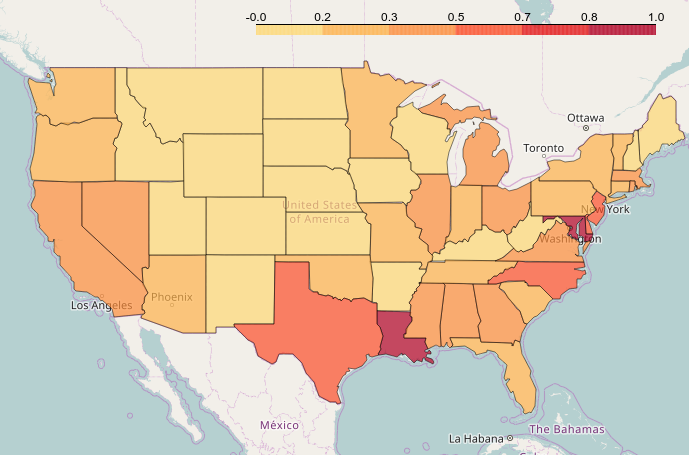}
    \end{subfigure}
    ~
    \begin{subfigure}[b]{0.24\textwidth}
    \centering
		\includegraphics[height=1.1in]{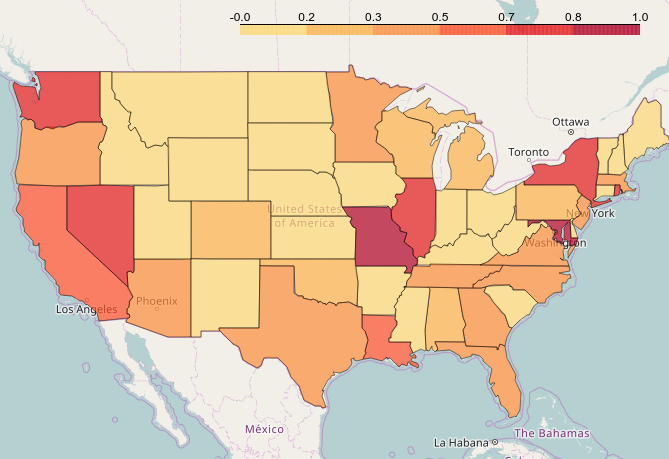}
    \end{subfigure}
\caption{Spatial distributions of statistics related to Black Lives Matter movement. (left) The frequency of the hashtag of BlackLivesMatter. (right) The frequency of violent protest images. The statistics are normalized by the number of users in each state.}
\label{fig:maps}
\end{figure}

Fig.~\ref{fig:maps} shows the spatial distributions of frequencies of the  \#BlackLIvesMatter hashtag and violent protest images in 2014-2016. 
Note as well that our classifier detects more violence in Missouri, Maryland, and New York.  Each state was the site of major protests after the deaths of Michael Brown (Ferguson, Missouri), Freddie Gray (Baltimore, Maryland) and Eric Garner (New York City, New York).  Investigating the temporal variation of the predictions of our classifier will provide more insight, as the classifier's violence predictions should spike during the protests, but the initial correlation between the classifier's prediction and our understanding of events is encouraging.

\begin{figure}
\vspace{-5pt}
    \centering
    \begin{subfigure}[b]{0.24\textwidth}
    \centering
		\includegraphics[height=1.1in]{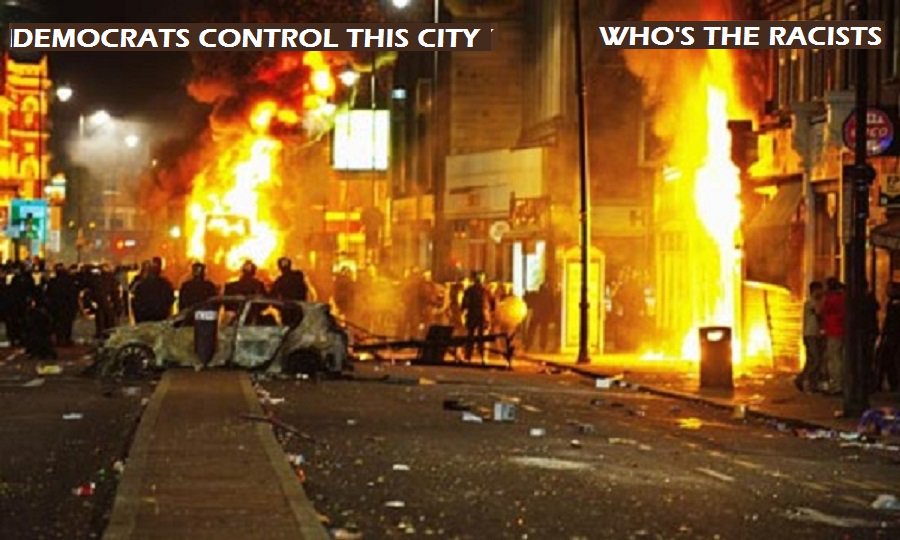}
        \caption{Text:-.599, Violence: .948}
        \label{fig:gull}
    \end{subfigure}
    ~
    \begin{subfigure}[b]{0.24\textwidth}
    \centering
		\includegraphics[height=1.1in]{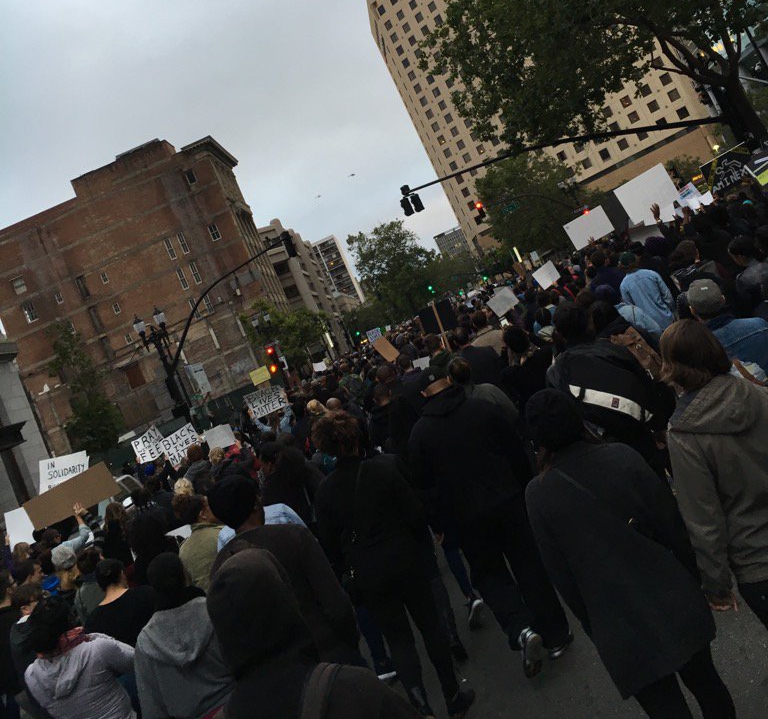}
        \caption{Text:-.946, Violence: .251}
        \label{fig:gull}
    \end{subfigure}
\caption{Two example tweets with text sentiment scores and image violence scores: (left) ``\#BlackLivesMatter okay \#WhiteLivesMatter okay but \#DemsLivesMatter UGH not so much when this happens.'' (right) ``protesting ignorance and fear that lead to  hate and violence. \#blacklivesmatter end \#policeshooting'' }
\label{fig:twttxt}
\end{figure}

\subsection{Multimodal Cues: Visual vs. Textual}
In order to examine the alignment between visual and textual cues in tweets, we measured the correlation between the text sentiment and the predicted image sentiments. The result is shown in Table~\ref{table:text_senti} and Fig.~\ref{fig:twttxt} shows two sample tweets. We used python's VADER (Valence Aware Dictionary and sEntiment Reasoner) package\cite{gilbert2014vader} which provides the sentiment measure from text. 12,055 tweets with protest images from Black Lives Matter and 10,566 tweets with protest images from Women's March were used to calculate the correlation. As expected, the visually inferred violence correlates negatively with positive text sentiment. However, the strength of the correlation is very weak, although they are statistically significant. This might be due the fact that tweet texts are very short or strong texts do not necessarily accompany strong images, and vice-versa. 

\begin{table}
\begin{center}
  \caption{Correlation coefficient between predicted text sentiment and perceived violence and image sentiment. ($p$-val $< 1.0\times10^{-10}$ for all variables.)}
  \label{table:text_senti}
  \begin{tabular}{ c | C{1cm}  C{1cm}  C{1cm}  C{1cm}  C{1cm}  }
              & Violent & Angry & Fearful & Sad & Happy \\ \hline
     Pearson's $\rho$ & -.080 & -.085 & -.088 & -.090  & .047 \\ \hline
  \end{tabular}
\end{center}
\end{table}

\section{Conclusion}
We have presented new approaches to estimate violence and protest dynamics from social media images. As our method is primarily based on visual analysis, it can generalize easier than textual analysis so long as visual language is more universal than spoken language. 

We constructed a large-scale novel dataset, UCLA Protest Image Dataset, which contains more than 10k protest images with their perceived violence values manually annotated. 
We will release the dataset with all the annotations collected for perceived violence and attributes. Using this data and a model trained on it, we have presented the results of our analysis on various past and on-going protest events in the world. 

Research in media studies and political science has suggested that the visual dimension of human communication can play a significant ``persuasive'' role in shaping public opinions \cite{Joo2014, soroka2016impact}. Our study demonstrates that the advances in computer vision and multimedia enable to systematically and automatically measure the impacts of visual media content to major social events in our society. 


Multimedia research has long investigated human emotion processing by computational approaches on large scale multimodal data. While its applications have reached out to a number of different disciplines, understanding social and political activities and their meanings and implications have been relatively overlooked. Therefore, our paper suggests a novel collaborative area of research between multimedia and political science.

\section{Acknowledgements}
This research is supported by UCLA TSG Program, ``Visual Big Data: Using Images to Understand Protests.'' We also acknowledge the support of NVIDIA Corporation for their donation of hardware used in this research.

\bibliographystyle{ACM-Reference-Format}
\bibliography{main} 

\end{document}